\documentclass[prl,twocolumn,superscriptaddress,showpacs,preprintnumbers,amsmath,amssymb,floatfix]{revtex4}

\usepackage{graphicx}
\usepackage{dcolumn}
\usepackage[latin1]{inputenc}
\usepackage[mathscr]{eucal}
\usepackage{amsmath}
\usepackage{amssymb}
\usepackage{epsfig}
\usepackage{sidecap}
\usepackage{bm}
\usepackage{bbm}
\usepackage{hyperref}

\newcommand{\aho}{a_{\scriptscriptstyle{\mathrm{ho}}}^{}}
\newcommand{\mf}{m_{\scriptscriptstyle{F}}^{}}
\newcommand{\br}{\bm{r}}
\newcommand{\tr}{\mathrm{tr}}
\newcommand{\sgx}{\sigma^x_{}}
\newcommand{\sgy}{\sigma^y_{}}
\newcommand{\sgz}{\sigma^z_{}}
\newcommand{\vtr}{V_{\scriptstyle{\mathrm{ext}}}}

\begin{document}
\preprint{ }

\title{Non-Abelian magnetic monopole in a Bose-Einstein condensate}

\author{Ville Pietil\"a}
\affiliation{Department of Applied Physics/COMP, Helsinki
  University of Technology P.~O.~Box 5100, FI-02015 TKK, Finland}
\affiliation{Australian Research Council, Centre of Excellence for Quantum Computer Technology,
The University of New South Wales, Sydney 2052, Australia}
\author{Mikko M\"ott\"onen}
\affiliation{Department of Applied Physics/COMP, Helsinki University of Technology P.~O.~Box 5100, FI-02015 TKK, Finland}
\affiliation{Australian Research Council, Centre of Excellence for Quantum Computer Technology,
The University of New South Wales, Sydney 2052, Australia}
\affiliation{Low Temperature
Laboratory, Helsinki University of Technology, P.~O.~Box 3500,
FI-02015 TKK, Finland}

\begin{abstract}
Recently, an effective non-Abelian magnetic field with a topology of a
monopole was shown to emerge from the adiabatic motion of multilevel atoms
in spatially varying laser fields [J.~Ruseckas {\it et al.},
Phys.~Rev.~Lett.~{\bf 95}, 010404 (2005)]. We study this monopole in a
Bose-Einstein condensate (BEC) of degenerate dressed states and 
find that the topological charge of the pseudospin cancels the monopole 
charge resulting in a vanishing gauge invariant charge.
As a function of the laser wavelength, 
different stationary states are classified in terms of their effect to the 
monopole part of the magnetic field and a cross-over to vortex ground state is 
observed. 
\end{abstract}


\maketitle

{\it Introduction.}~--- Existence of an isolated point source of magnetic 
field, that is, a
magnetic monopole, has been an intriguing question ever since Dirac
proposed that the existence of a magnetic monopole leads in a natural way
to the quantization of electric charge~\cite{Dirac1931}. The magnetic 
monopole considered by Dirac is intrinsically a singular object with
an string-like singularity attached to it. In the seminal work of 't Hooft
and Polyakov, non-singular monopole configurations were shown to arise
in the context of non-Abelian gauge theories~\cite{tHooft1974,Polyakov1974}. 
Unfortunately, these kind of 
monopoles are expected to be extremely heavy which renders their
experimental detection difficult. Despite ingenious
theoretical and experimental work in the fields of high energy
physics, cosmology, and condensed matter physics, convincing
evidence of the existence of stable magnetic monopoles is still 
lacking~\cite{Milton2006}. 

Advances in trapping and manipulating 
degenerate quantum gases have revealed the potential of the cold atom
systems to serve as quantum simulators for ideas beyond the usual 
condensed matter phenomena. In particular, it has been  
proposed that non-Abelian gauge potentials are realized in the 
effective description of atoms with degenerate 
internal degrees of freedom coupled to spatially varying laser 
fields~\cite{Osterloh2005,Ruseckas2005,Lu2007,Liu2007a}. 
Based on the earlier work by Wilczek and Zee~\cite{Wilczek1984},
Ruseckas {\it et al.}~\cite{Ruseckas2005} have shown that
non-Abelian gauge potentials describe the off-diagonal couplings between
the degenerate dressed states in atoms with multiple degenerate
internal states. As a specific example, a $U(2)$ gauge potential
corresponding to a non-Abelian magnetic monopole was
constructed~\cite{Ruseckas2005}. 

Previous monopole studies in Bose-Einstein condensates (BECs) 
have concentrated on systems where the monopole state occurs without
any gauge potential and stems essentially from the non-linear
interactions between atoms in different hyperfine spin 
states~\cite{Stoof2001}. This renders such monopoles typically energetically 
unfavorable~\cite{Stoof2001}. 
On the other hand, atom-laser interaction induced Abelian magnetic
monopoles and non-Abelian monopoles in optical lattices have been investigated 
recently~\cite{Zhang2005}.
In this Letter, we consider an isolated non-Abelian 
magnetic monopole in a Bose-Einstein
condensate, which is similar to the monopoles in the Yang-Mills-Higgs 
model~\cite{Arafune1975}. We study the lowest energy states of the system and 
show that they can be classified according to the topological charge of the 
pseudospin texture generated by gauge transformations. On the other hand, we 
find the gauge invariant total charge of the system to be always zero.

{\it The model.}~--- Consider a four-level tripod scheme,
in which degenerate atomic states $\{|1\rangle,|2\rangle,|3\rangle\}$ are 
excited to a common virtual state $|0\rangle$, see~\cite{Ruseckas2005}. 
In the rotating wave approximation the atom-light interaction Hamiltonian
has two degenerate dark states $|\chi_1\rangle$ and $|\chi_2^{}\rangle$ 
corresponding to zero eigenvalue, and two bright states $|\chi_3^{}\rangle$  
and $|\chi_4^{}\rangle$ with a non-zero projection to the radiatively 
decaying excited level $|0\rangle$. 
Expressing the full quantum state $|\Phi\rangle =
\sum_{k=0}^{3}\zeta_k(\br)|k \rangle$ 
in the basis of the dressed states as $|\Phi\rangle =
\sum_{m=1}^{4}\psi_m(\br)|\chi_m^{}(\br) \rangle$ gives rise to a unitary
transformation $\mathcal{U}_{mk}(\br) = \langle\chi_m^{}(\br)|k\rangle$
 between vectors $\bm{\Psi} =
(\psi_1\,\cdots\,\psi_4)^T$ and $\bm{\zeta} =
(\zeta_0\,\cdots\,\zeta_3)^T$ as $\psi_m(\br) =
\sum_{k=0}^3\mathcal{U}_{mk}(\br)\zeta_k(\br)$. 
If transitions from the degenerate dark states to bright states are
neglected one obtains an effective single-particle Hamiltonian for the dark  
states 
\begin{equation}
\hat{H}_1 =
\int\mathrm{d}\br\,\left[\frac{\hbar^2}{2m}(D_{\mu}\hat{\bm{\psi}})^{\dagger}(D_{\mu}\hat{\bm{\psi}})
+ \hat{\bm{\psi}}^{\dagger}(\vtr + \Phi)\hat{\bm{\psi}}\right],
\end{equation}
where $D_{\mu} = \partial_\mu - iA_{\mu}$  and $\hat{\bm{\psi}} = 
(\hat{\psi}_1,\hat{\psi}_2)^T$, see Ref.~\cite{Ruseckas2005}. The
Berry connection $\bm{A} = A_{\mu}\bm{e}_{\mu}$ and the scalar
potential $\Phi$ transform as 
$\bm{A}\rightarrow U\bm{A}U^{\dagger} - i(\nabla U)U^{\dagger}$ and 
$\Phi\rightarrow U\Phi U^{\dagger}$ under a local change of basis given by 
a unitary matrix $U(\br)$. Furthermore, we denote $A_{\mu} = A_{\mu}^{\alpha}
\sigma^{\alpha}_{}$ and 
$\sigma_{}^{\alpha} = (\mathbbm{1},\sgx,\sgy,\sgz)$~\footnote{
\label{fnote} The
  Greek indices $\mu,\nu$... refer to the spatial 
coordinates and $\alpha$ runs over values $0,\cdots3$. Indices
$a,b$... refer to the Pauli matrices $\bm{\sigma} =
(\sgx,\sgy,\sgz)$. 
Summation over repeated indices is implied.}. The external potential
that confines the atoms is given by $\vtr$. 

The non-Abelian magnetic monopole was shown to
arise from two circularly polarized beams and a linearly polarized
beam yielding~\cite{Ruseckas2005} 

\begin{align}
\label{vector}
&\bm{A} = -\frac{\cos\vartheta}{r\sin\vartheta}\bm{e}_{\varphi}^{}\sgx
+ \frac{1}{2}(k\bm{e}_z^{}-k'\bm{e}_x^{}) \notag \\ 
&\hspace{3cm}\times[(1+\cos^2\vartheta)\mathbbm{1}+\sin^2\vartheta\,\sgz], \\
\label{scalar}
&\Phi = \frac{\hbar^2}{2m}\bigg[\frac{1}{r^2}\mathbbm{1} 
-\frac{k'}{2r}\sin\varphi\sin 2\vartheta\,\sgx \notag \\
&\hspace{3cm}+ \frac{k^2 +
  k'^2}{8}\sin^2 2\vartheta (\mathbbm{1}-\sgz)\bigg],
\end{align}  
where $(r,\varphi,\vartheta)$ denote the spherical coordinates.
As pointed out in~\cite{Ruseckas2005}, the effective magnetic 
field $B_{\mu} =
\frac{1}{2}\varepsilon_{\mu\nu\lambda}^{}F_{\nu\lambda}^{\alpha}\sigma_{}^{\alpha}$ is of the form $\bm{B} =
1/r^2\,\bm{e}_r\sgx+\cdots$ in which the omitted  ``non-monopole''
terms do not generate a net flux through a closed surface enclosing
the monopole. The field
strength tensor $F_{\mu\nu}^{\alpha}$ is given by $F_{\mu\nu}^{a} =
  \partial_{\mu}A_{\nu}^{a}-\partial_{\nu}A_{\mu}^{a} + 2\varepsilon^{abc}_{}A_{\mu}^{b}A_{\nu}^{c}$ and $F_{\mu\nu}^{0} =
  \partial_{\mu}A_{\nu}^{0}-\partial_{\nu}A_{\mu}^{0}$, see~[19]. 

Let us consider the atom-laser and the atom-atom interactions in the 
original basis $\{|k\rangle\},\,k=0,\cdots,3$.  
For a system consisting of, e.g.,~$^{87}$Rb atoms with 
hyperfine spin $F$, the set of degenerate internal states 
$\{|1\rangle,|2\rangle,|3\rangle\}$ can be taken to be the Zeeman sublevels 
$|F=1,\mf\rangle$, $\mf = 1,-1,0$ at the $5S_{1/2}^{}$ multiplet, and the 
excited state e.g.~$|F=0,\mf=0\rangle$ at the $5P_{3/2}$ multiplet. 
For $^{87}$Rb the scattering lengths in the channels corresponding to 
total hyperfine spin $f=0$ and $f=2$ of the two scattering particles are almost 
equal~\cite{vanKampen2002}, and hence we neglect the spin-dependent
part of interaction between states $|F=1,\mf\rangle$, see~\cite{Ho1998}. 
Furthermore, similarly to Ref.~\cite{Liu2007a} we 
assume that the population of the two bright states $|\chi_3\rangle$ and   
$|\chi_4\rangle$ is vanishingly small compared to the population of the
dark states $|\chi_1\rangle$ and $|\chi_2\rangle$. This is, in fact, 
a non-trivial assumption since for a three-level 
$\Lambda$-system it has been shown that the dark
state can be unstable under small deviations, leading to non-vanishing
population of the bright state~\cite{Itin2007}. 
Similar analysis for the tripod system
considered here has yet to be performed. Using the 
unitary transformation $\mathcal{U}_{mk}(\br)$ and the Rabi
frequencies~\cite{Ruseckas2005} $\Omega_{1,2}(\bm{r}) =  
\Omega_0\frac{\rho}{\sqrt{2}R}e^{i(kz\mp\varphi)}_{}$, 
$\Omega_3(\bm{r}) = \Omega_0\frac{z}{R}e^{ik'x}_{}$ corresponding to the 
monopole field, one obtains an effective Hamiltonian 
$\hat{H}_{\scriptscriptstyle AA} = \frac{c_0^{}}{2}\int\mathrm{d}\br\,(\hat{\bm{\psi}}^{\dagger}\hat{\bm{\psi}})^2
$ describing the interactions in the dark state manifold. 

In the absence of light induced gauge potentials 
Bose-Einstein condensation takes place at low enough
temperatures. %On the other hand, existence of artificial gauge
%potentials can alter the formation of BEC. For the 
%tripod scheme considered here, effective light-induced spin-orbit
%interactions have been shown to give rise to a BEC
%in which the bosons condense into an entangled quantum 
%state~\cite{Stanescu2007}. 
%With this caveat in mind we seek further insight to the condensate
%states by assuming a single macroscopically populated mode. 
%This approach cannot directly treat effects such as condensate
%fragmentation, but even in the absence of a coherent condensate the
%mean field states can be used to construct approximations for the true ground
%state as well as to interpret the possible outcomes of a particular
%experimental realization~\cite{Castin2001}. 
On the other hand, the existence of artificial gauge potentials can
alter the formation of BEC, leading to fragmentation~\cite{Stanescu2007}. 
Although strictly speaking the mean field approach used below cannot
treat effects such as fragmentation, even in the absence of a
true condensate the mean field states can be used to construct
approximations for the true ground state as well as to interpret
the possible outcomes of a particular experimental 
realization~\cite{Castin2001}.  
Replacing $\hat{\bm{\psi}}$ by $\langle\hat{\bm{\psi}}\rangle = \bm{\psi}$ we
obtain a mean field Hamiltonian     
\begin{equation}
\label{hami}
\mathcal{H} =
\int\mathrm{d}\br\,\bigg[\frac{\hbar^2}{2m}(D_{\mu}\bm{\psi})^{\dagger}(D_{\mu}\bm{\psi})
+ \bm{\psi}^{\dagger}(\vtr + \Phi)\bm{\psi} + \frac{c_0^{}}{2}|\bm{\psi}|^4\bigg].
\end{equation} 
The atom-laser interaction
is diagonal in the basis consisting of the dressed states
$\{|\chi(\br)\rangle\}$. The dark states correspond to a zero 
eigenvalue and hence the atom-laser interaction does not appear 
in Eq.~\eqref{hami}. For optically confined spin-$1$ bosons such as 
$^{87}$Rb, the external trapping potential $\vtr$ can be treated as a
scalar function~\cite{Ho1998} and the model in Eq.~\eqref{hami} is  
$U(2)$ gauge invariant. 

{\it Gauge invariance.}~--- In the Abelian case, the magnetic monopole 
is accompanied with singular filament, the Dirac string, that extends
outwards from the monopole. For non-Abelian monopoles 
it has been pointed out that the Dirac string can be removed by a
suitable gauge transformation~\cite{Arafune1975}. Similar line of reasoning
applies also here and by applying a gauge transformation
$U_0(\br) = e^{i\varphi\sgx}_{}e^{-i\vartheta\sgy/2}_{}$ one can  
remove the divergent part of $\bm{A}$. 
The transformed vector potential is given by  
\begin{align}
\label{nonsing}
\bm{A}' &= \frac{1}{r}\bm{e}_{\varphi}^{}b^a_{}\sigma^a_{} -
\frac{1}{2r}\bm{e}_{\vartheta}^{}[\cos(2\varphi)\sgy - \sin(2\varphi)\sgz] \notag \\
&+
\frac{1}{2}(k\bm{e}_z^{}-k'\bm{e}_x^{})[(1+\cos^2\vartheta)\mathbbm{1} +\sin^2\vartheta\,b^a_{}\sigma^a_{}],
\end{align}
with $\bm{b} =
(\sin\vartheta,\,\sin(2\varphi)\cos\vartheta,\,\cos(2\varphi)\cos\vartheta)$. 
The scalar potential transforms according to   
$\Phi' = U_0 \Phi U^{\dagger}_0$. It should be noted that $\bm{A}'$ is
not well-defined at the $z$ axis although the gauge transformation
removed the explicit divergence. The scalar potential $\Phi'$ on the other 
hand, remains well-defined at the $z$ axis.

The topological structure of the condensate can be analyzed by 
studying the pseudospin 
\begin{equation}
\label{pseudospin}
\bm{s}(\bm{r}) = \bm{\psi}^{\dagger}(\bm{r})\bm{\sigma}\bm{\psi}(\bm{r}).
\end{equation}
We note that $\bm{s}$ is not itself gauge invariant,
but transforms under any gauge
transformation $U$ as $s^{a}_{} \rightarrow u^{ab}_{}s^{b}_{}$ where 
$U^{\dagger}\sigma^{a}_{}U = u^{ab}_{}\sigma^{b}_{}$. An explicit computation 
of the matrix elements $u^{ab}_{}$ shows that the gauge transformation 
of the pseudospin $s^a_{}$ can be written as
\begin{equation}
\label{sgaugetransf}
s^{a}_{}\sigma_{}^{a} \rightarrow Us^{a}_{}\sigma_{}^{a} U^{\dagger}.
\end{equation}
We define the covariant derivative of the pseudospin by
\begin{equation}
\label{ds}
\mathcal{D}_{\mu} s^a_{} = \bm{\psi}^{\dagger}\sigma_{}^a D_{\mu}\bm{\psi} + 
\mathrm{c.c.} = \partial_{\mu} s^a_{} +
2\varepsilon^{abc}_{}A_{\mu}^b s_{}^c,
\end{equation}
from which we see that $\mathcal{D}_{\mu} s^a_{}$ transforms as 
$\mathcal{D}_{\mu} s^a_{} \rightarrow u^{ab}_{}\mathcal{D}_{\mu} s^b_{}$. 
Using identities $2|\bm{s}|^2 = \tr(s^a_{}\sigma^a_{}s^b_{}
\sigma^b_{})$, $2\varepsilon^{abc}_{}s^a_{}\mathcal{D}_{\mu} s^b_{}
\mathcal{D}_{\nu} s^c_{} = -i\,\tr[s^a_{}\sigma^a_{}(\mathcal{D}_{\mu} s^b_{})
\sigma^b_{}(\mathcal{D}_{\nu} s^c_{})\sigma^c_{}]$, and 
Eq.~\eqref{sgaugetransf} one observes that 
\begin {equation}
\label{emtensor}
G_{\mu\nu} = \frac{1}{|\bm{s}|}F_{\mu\nu}^a s^a_{} - \frac{1}{2|\bm{s}|^3}
\varepsilon^{abc}s^a_{}\mathcal{D}_{\mu} s^b_{}\mathcal{D}_{\nu} s^c_{},
\end{equation}
is invariant under $U(2)$ gauge transformations of $\bm{\psi}$ and
$A_{\mu}$. Tensor $G_{\mu\nu}$ is the ``electromagnetic'' tensor introduced 
by 't Hooft in the studies of monopoles in unified gauge 
theories~\cite{tHooft1974}, although here $\bm{s}$ is not the 
fundamental variable but defined by Eq.~\eqref{pseudospin}. 

Using a unit vector field $\hat{\bm{n}} = \bm{s}/|\bm{s}|$ one can
write $G_{\mu\nu}$ in 
the form~\cite{Arafune1975} $G_{\mu\nu} = M_{\mu\nu} - H_{\mu\nu}$
where 
\begin{align}
\label{magnetic}
M_{\mu\nu} &= \partial_{\mu}(\hat{n}^a_{}A_{\nu}^a) - \partial_{\nu}(\hat{n}^a_{}A_{\mu}^a), \\ 
\label{pspin}
H_{\mu\nu} &= \frac{1}{2}\varepsilon^{abc}_{}\hat{n}^a_{}\partial_{\mu}\hat{n}^b_{}\partial_{\nu}\hat{n}^c_{}.
\end{align}
In particular, Eqs.~\eqref{magnetic}--\eqref{pspin} imply that 
$G_{\mu\nu}$ is linear with respect to the gauge potential
$A_{\mu}^a$. This enables us to study separately the terms 
in the gauge potential which are responsible for the monopole 
and consider the terms depending on $k$ or $k'$ as a background field 
for the monopole configuration, see Eq.~\eqref{nonsing}. Using the 
tensor $G_{\mu\nu}$ we define magnetic charge density   
$J = \varepsilon_{\mu\nu\lambda}^{}\partial_{\mu}G_{\nu\lambda}$, and 
with the Gauss' theorem the total magnetic charge can be written 
as~\cite{Arafune1975}
\begin{equation}
\label{topocharge}
Q = \frac{1}{8\pi}\int_{S^{\infty}}\mathrm{d} \tau_{\mu}\,\varepsilon_{\mu\nu\lambda}^{}[M_{\nu\lambda} - H_{\nu\lambda}] = Q_M^{} - Q_S^{}.
\end{equation} 
where $\mathrm{d} \tau_{\mu}$ is the surface element on
$S^{\infty}$. The factor of $2$ in the derivative of $s^a$
[Eq.~\eqref{ds}] implies that $Q_S^{}$ is quantized in the units of
$\frac{1}{2}$, that is, $Q_S^{}$ is $\frac{1}{2}$ times the winding
number of the unit vector field $\hat{\bm{n}}$, see~\cite{Arafune1975}.  

The gauge transformation $U_0$ can be decomposed as $U_0 = U_2 U_1$
where and $U_2 =
e^{i\varphi\sgx}e^{i\vartheta\sgy/2}e^{-i\varphi\sgx}$ and $U_1 = e^{i\varphi\sgx}$.
In particular, transformation with $U_1$ brings the monopole part of
$\bm{A}$ to the form which has the Dirac string at negative $z$ axis
and is thus analogous to the Dirac description of the Abelian 
monopole~\cite{Dirac1931}. The meaning of the different gauges can be 
understood qualitatively by considering the following ansatz 
$
\bm{\psi}_{\pm}(\br) =\phi(\br)\bm{\zeta}_{\pm}^{}\,\,\,\,
\mathrm{with}\,\,\bm{\zeta}_{\pm}^{} = (1/\sqrt{2},\mp 1/\sqrt{2})^T
$
for the condensate order parameter in the non-singular gauge  
$\mathcal{G}_2 = (\bm{A}',\Phi')$. For this simple trial wavefunction 
one obtains the following diagram
\begin{equation}
\label{anzats}
\begin{array}{c}
\mathcal{G}_2 \\
Q_M^{(2)} = 0 \\
Q_S^{(2)} = 0
\end{array}\,\,
\xrightarrow{U_2^{\dagger}}\,\,
\begin{array}{c}
\mathcal{G}_1 \\
Q_M^{(1)} = \pm 1 \\
Q_S^{(1)} = \pm 1
\end{array}\,\,
\xrightarrow{U_1^{\dagger}}\,\,
\begin{array}{c}
\mathcal{G}_0 \\
Q_M^{(0)} = 0 \\
Q_S^{(0)} = 0,
\end{array}
\end{equation} where 
$\mathcal{G}_0 = (\bm{A},\Phi)$ is the original gauge given by 
Eqs.~\eqref{vector} and~\eqref{scalar} and $\mathcal{G}_1$ is obtained
from $\mathcal{G}_0$ with $U_1$. In the next section we will use the 
intermediate gauge $\mathcal{G}_1$ to classify different stationary states.

\begin{SCfigure*}
\centering 
\includegraphics[width=0.7\textwidth]{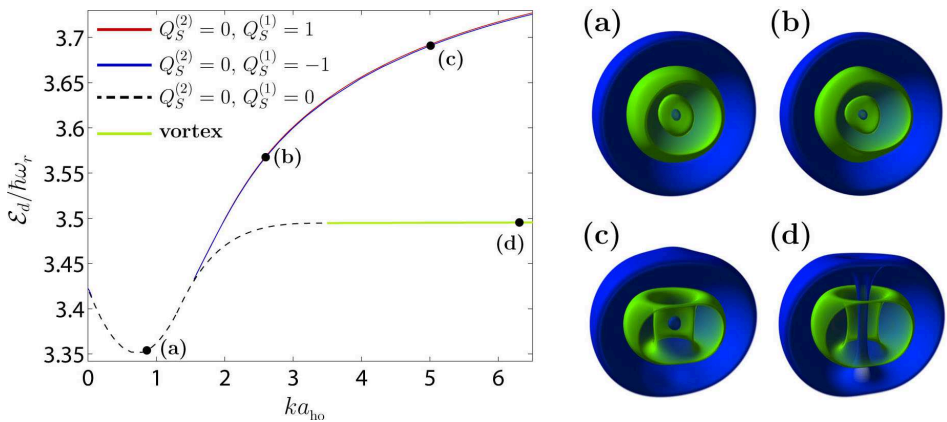}   
\caption{\label{denergy} (Color online) Left panel: mean field energies 
corresponding to the solutions of the Gross-Pitaevskii equation. States 
with $Q_S^{(1)} = \pm 1$ are practically indistinguishable in the
plot. Right panel: isosurfaces of the particle
densities for $k\aho = 0.86$ (a), $k\aho = 2.6$ (b), $k\aho = 5.0$
(c), and $k\aho = 6.3$ (d). The corresponding points are marked with 
\textbullet~in the energy diagram. In panel (d) the ground state has a 
vortex. The particle density increases from dark (blue) to light (green) 
color.}  
\end{SCfigure*} 

{\it Numerical analysis.}~--- To study the lowest energy states we
search numerically solutions to the field equations corresponding to the
Hamiltonian in Eq.~\eqref{hami}. We discretize the Hamiltonian 
using the scheme previously utilized in the lattice gauge theory
studies of the non-Abelian Higgs model~\cite{Fradkin1979}. We replace
$(D_{\mu}\bm{\psi})^{\dagger}(D_{\mu}\bm{\psi})$ by 
$
C_{\mu , X} = (\bm{\psi}^{}_{X+\hat{\mu}} - U_{\mu , X}^{}\bm{\psi}^{}_{X})^{\dagger}(\bm{\psi}^{}_{X+\hat{\mu}} - U_{\mu , X}^{}\bm{\psi}^{}_{X})/a^2,
$
where $X = (u,v,w) \in \mathbbm{N}^3$ denotes one lattice site, $U_{\mu , X}^{} =  e^{iaA'_{\mu , X}}_{}$ is the link variable,
and $\hat{\mu}$ is one lattice
unit in the direction $\mu$ in the lattice. When the lattice
constant $a$ tends to zero, $C_{\mu , X}$  reduces to
$(D_{\mu}\bm{\psi})^{\dagger}(D_{\mu}\bm{\psi})$. 
The rest of the Hamiltonian is readily discretized yielding the mean field 
energy 
\begin{equation*}
\mathcal{E}_d = \frac{\hbar^2}{2m}\sum_{X,\mu}C_{\mu,X}^{} + \sum_X [\bm{\psi}_X^{\dagger}(V_{\mathrm{ext}, X} + \Phi_X)\bm{\psi}^{}_{X} + \frac{c_0^{}}{2}|\bm{\psi}^{}_{X}|^4]
\end{equation*}
Minimizing the free energy  
$\mathcal{F}_d = \mathcal{E}_d - \mu_{\mathrm{ch}}^{}\sum_X |\bm{\psi}^{}_{X}|^2$  
gives the discrete version of the time-independent 
Gross--Pitaevskii~(GP) equation 
\begin{align}
\label{gpe}
&-\frac{\hbar^2}{2ma^2}\sum_{\mu = x,y,z}(U_{\mu ,
  X}^{\dagger}\bm{\psi}^{}_{X+\hat{\mu}} -2\bm{\psi}^{}_{X} + U_{\mu ,
  X-\hat{\mu}}^{}\bm{\psi}^{}_{X-\hat{\mu}}) \notag \\
&+(V_{\mathrm{ext} , X} + \Phi_X)\bm{\psi}^{}_{X} +
c_0|\bm{\psi}^{}_{X}|^2\bm{\psi}^{}_{X} = \mu_{\mathrm{ch}}^{}\bm{\psi}^{}_{X}.
\end{align}
The chemical potential guaranteeing the conservation of the particle 
number is denoted by $\mu_{\mathrm{ch}}^{}$.

For the numerical calculation we assume the external potential 
of the form $\vtr = m\omega_r^{2} r^2/2$ and for simplicity 
that the wave vectors of the laser fields are equal and $k = k' > 0$. 
The spatial variables are scaled with the harmonic oscillator length 
$\aho = \sqrt{\hbar/m \omega_r^{}}$. The strength of the inter-particle 
interaction is taken 
to be $c_0^{}N/\hbar\omega_r^{} = 100$. We solve the 
GP~equation~\eqref{gpe} iteratively using the successive 
over-relaxation scheme (SOR) with periodic 
boundary conditions in a grid of $121^3$ points. Since the vector 
potential in Eq.~\eqref{nonsing} is not well defined at the $z$ axis
we shift the grid in $x$-direction by a positive constant $\eta$ taken 
to be $\eta=a/10$. The results are independent of the choice of
$\eta$. The mean field energy of different stationary states is 
shown in Fig.~\ref{denergy} as a function of $k\aho$. The 
calculations are carried out in the non-singular gauge 
$\mathcal{G}_2 = (\bm{A}',\Phi')$ for which the corresponding charges
are denoted by $Q^{(2)}_S$ and $Q^{(2)}_M$. 

All stationary states have $Q^{(2)}_S = Q^{(2)}_M = 0$ for 
$k\aho \lesssim 3.5$. For $k\aho\gtrsim 3.5$ the 
ground state contains a vortex and the classification in terms of 
Eq.~\eqref{topocharge} is no longer valid.
For $k\aho \lesssim 0.055$ there is a pair of solutions which are
degenerate within the numerical accuracy. These solutions behave under the 
gauge transformation $U_2^{\dagger}$ such that 
$Q^{(2)}_S = 0 \rightarrow Q_S^{(1)} = \pm 1$, where the charges in the 
intermediate gauge $\mathcal{G}_1$ are denoted by $Q_S^{(1)}$ and $Q_M^{(1)}$.
For $k\aho \gtrsim 2.1$ the  
pair appears again as excited states with $Q_S^{(1)} = 1$ state having slightly 
larger energy. The difference in the energies between 
these two states is, however, very small and cannot be easily distinguished 
from Fig.~1. In the original gauge $\mathcal{G}_0$,  
we find $Q^{(0)}_S = 0$ for all found states, similar to Eq.~\eqref{anzats}. 
For  $k\aho \lesssim 3.5$ the lowest energy state is always the state with 
$Q_S^{(1)} = 0$. 

In the gauge $\mathcal{G}_0$, the single-particle operator 
$\hat{h}_0$ of the mean field Hamiltonian $\mathcal{H}=\int\mathrm{d}\br\,[\bm{\psi}^{\dagger}\hat{h}_0\bm{\psi}+\frac{c_0^{}}{2}|\bm{\psi}|^4]$ has the symmetry 
$\hat{h}_0(\varphi) = \sgz\hat{h}_0(-\varphi)\sgz$. The two-fold symmetry 
explains the degeneracy of $Q_S^{(1)} = \pm 1$ states for small $k\aho$ and 
suggests that the small energy difference between these two states for 
large $k\aho$ can be due to the discrete lattice that breaks the 
symmetry with respect to rotations about the $z$ axis. For small $k\aho$, the 
energies of $Q_S^{(1)} = \pm 1$ and $Q_S^{(1)} = 0$ states almost coincide   
and they are indistinguishable in Fig.~1. The transition from the 
ground state with $Q^{(1)}_S = 0$ to the vortex ground state occurs
when initially far apart separated pair of coreless vortices 
starts to become more and more tightly bound with increasing $k\aho$
eventually forming a singular vortex. Once the vortex state is formed, its 
energy seems to depend only weakly on $k\aho$.

Within the mean field theory employed here, we cannot say anything definite 
about the possible fragmentation of the condensate. If the possible 
fragmentation is in this case related to restoring the broken 
symmetry~\cite{Castin2001}, the true ground state could correspond to a 
fragmented condensate due to the two-fold symmetry of the mean field 
Hamiltonian. Numerically, however, we seem to find only a non-degenerate ground 
state. To study whether the BEC in a monopole field is fragmented, 
one needs to perform a calculation analogous to the one in 
Ref.~\cite{Stanescu2007}.

Assuming that the underlying system consists of $^{87}$Rb
atoms, the two obvious gauge invariant observables are 
the particle density $\varrho$ and the hyperfine spin of the
constituent atoms. Both of these quantities can be imaged accurately with the
state-of-the-art techniques. In the original gauge $\mathcal{G}_0$,
the pseudospin $\bm{s}$ can be expressed in term these two quantities
which in principle enables reconstruction of $\bm{s}$. Thus it should be 
experimentally possible to distinguish between states with $Q^{(1)}_S =
\pm 1$ and $Q^{(1)}_S = 0$  by performing a local change of basis using $U_1$.  
In the current experiments with $^{87}$Rb, one typically has $1.5 \lesssim k\aho
\lesssim 10$ for which the ground
state either has $Q^{(1)}_S = 0$ or contains a vortex. For very 
tight traps with $\omega_r^{}$ of order of kilohertz, the point where 
$Q^{(1)}_S =\pm 1$ and $Q^{(1)}_S = 0$ states start to coexist can be reached. 
By changing the trap parameters it is then possible to tune the value of 
$k\aho$ and  explore different regions of the mean field phase diagram in 
Fig.~1. In particular, changing $\omega_r^{}$ non-adiabatically could result 
in the $Q^{(1)}_S = \pm 1$ state. 

In conclusion, we have studied a non-Abelian monopole in a 
Bose-Einstein condensate and showed that the system can be described 
with an effective $U(2)$ gauge invariant model. We identified a gauge 
invariant charge characterizing the system and classified different 
stationary states using this charge. Numerical calculations showed 
that the existence of a monopole in the non-Abelian gauge potential gives 
rise to a pseudospin texture with a topological 
charge that cancels the monopole charge.  

The authors would like to thank M.~Nakahara for 
discussions and Jenny and Antti Wihuri
Foundation, Emil Aaltonen Foundation, and the Academy of Finland for 
financial support. CSC, the Finnish IT center for science, is acknowledged 
for computational resources.

\bibliography{manu}

\end{document}